# Component periods of non-principal-axis rotation and their manifestations in the lightcurves of asteroids and bare cometary nuclei

**Nalin H. Samarasinha and Beatrice E.A. Mueller**

Planetary Science Institute
1700 E Ft Lowell Road, Tucson, AZ 85719, USA

Pages: 31
Figures: 7
Tables: 2

Corresponding Author: Nalin H. Samarasinha (nalin@psi.edu)

# Abstract

Lightcurve observations of asteroids and bare cometary nuclei are the most widely used observational tool to derive the rotational parameters. Therefore, an in-depth understanding of how component periods of dynamically excited non-principal axis (NPA) rotators manifest in lightcurves is a crucial step in this process. We investigated this with the help of numerically generated lightcurves of NPA rotators with component periods known *a priori*. The component periods of NPA rotation were defined in terms of two widely used yet complementary conventions. We derive the relationships correlating the component rotation periods in the two conventions. These relationships were then used to interpret the periodicity signatures present in the simulated lightcurves and rationalize them in either convention.

# 1. Introduction

Dynamically, the most stable rotational state of an object is the one that requires the least amount of kinetic energy of rotation for a given total rotational angular momentum. This corresponds to the rotation about its axis of maximum moment of inertia (i.e., about its short principal axis). A small body of the solar system such as an asteroid or a comet nucleus in this least energy rotational state could become a rotationally excited object due to external torques acting on it (e.g., torques due to outgassing in the case of a comet nucleus or due to YORP torques in the case of an asteroid) or as a consequence of changes to its principal moments of inertia (e.g., due to significant mass loss in the case of a comet nucleus or large impacts in the case of an asteroid). Such rotationally excited bodies are in general (i.e., except in the limiting cases where the rotation is either about the axis of minimum moment of inertia or the axis of intermediate moment of inertia[1]) in rotational states where the rotation is not about a principal axis and therefore are known as non-principal-axis rotational states (NPA rotational states). NPA rotational states are also known in the literature as complex rotational states or objects in "tumbling motion". The likelihood of small bodies being in NPA rotational states is discussed in the literature based on theoretical considerations (e.g., Prendergast 1958, Burns and Safronov 1973, Jewitt 1997). With the discovery of NPA rotation among small bodies beginning with comet 1P/Halley (e.g., Belton et al. 1991, Samarasinha and A'Hearn 1991) and asteroid 4179 Toutatis (Hudson and Ostro 1995), it was apparent that some small bodies could indeed be in dynamically excited NPA rotational states. For such an object, to be in an NPA state, the damping timescale due to mechanical friction must be larger than either the time elapsed since the last excitation event (e.g., due to an asteroidal impact or a splitting event) or the rotational excitation timescale (e.g., rotational excitation timescale due to outgassing torques on a comet nucleus).

NPA rotation is discussed in detail in standard dynamics textbooks (e.g., Ames and Murnaghan 1929, Landau and Lifshitz 1976). In addition, it has been discussed in terms of small body rotational dynamics (e.g., Samarasinha and A'Hearn 1991, Kaasalainen 2001). However, for the purpose of simplicity, in many textbooks, the detailed discussion on the NPA rotation is limited to symmetric objects (i.e., objects with an axis of symmetry). Furthermore, in many textbooks, the discussion is geared towards explaining the NPA rotation as pertaining to the Earth. As a consequence, the NPA rotation is defined in a way to easily explain the temporal behavior of the short principal axis of the object (hereafter, we will use the term "S-convention" to refer to this; also see Appendix A). However, in describing the rotational motion of an elongated object, a different convention may be preferred, so one can easily explain the temporal behavior of the long axis. The choice of convention used in Samarasinha and A'Hearn (1991) (hereafter SA91) follows this, as there the primary objective was to explain the NPA rotation of the elongated nucleus of comet 1P/Halley. We will use the term "L-convention" to refer to this. From an observer's point of view, differentiating and identifying the long axis of an elongated body is much easier than identifying either the short or the intermediate axes as the latter dimensions are closer to each other. However, there are a number of publications (e.g., Kaasalainen 2001) that describe the component rotational periods in the S-convention, which is best suited to explain the behavior of the short principal axis.

---

[1] Rotation about the axis of intermediate moment of inertia is dynamically unstable.

The component rotational periods associated with the NPA rotation manifest in the lightcurves of asteroids and bare cometary nuclei[2] in the form of signatures of lightcurve periodicities that one can detect by analysis of time series such as by Fourier analyzing a lightcurve. A deeper and precise understanding of the signatures of these periodicities present in the lightcurve of an NPA rotator is possible if we are familiar with the relationships between the rotational periods with respect to these two conventions. This would also facilitate our understanding, at least to an extent, why those signatures but not others exist in the rotational lightcurves to begin with. For this purpose, we will make use of the results from numerically simulated (i.e., synthetic) lightcurves with known rotational states and component rotational periods.

In Section 2 of this paper, we compare and contrast the properties of the NPA rotators in both these conventions. Then we introduce the basic formulae that relate the component rotational periods defined in terms of the two conventions. Section 3 analyzes the periodicity signatures of a number of numerically simulated lightcurves and those periodicities will be rationalized in terms of the component periods of rotation defined in either convention. Section 4 presents the conclusions of this paper. Appendix A of this paper is analogous to Appendix A of SA91 but presented in terms of the S-convention and the relevant properties are discussed extensively in Section 2.

## 2. Component Rotational Periods of the NPA Rotation in the L-convention and S-convention

The component periods associated with the NPA rotation are generally expressed in terms of Euler angles (e.g., Landau and Lifshitz 1976). Such component periods can be associated with component motions of the NPA rotation, which makes sense from an external observer's point of view. There are three associated component periods: $P_\psi$, $P_\theta$, and $P_\phi$. However, $P_\psi$ and $P_\theta$ are coupled to each other (see Section 2.1) and therefore we have only two independent component periods: $P_\psi$ and $P_\phi$. So, how do component rotational periods $P_\psi$ and $P_\phi$ defined in the L-convention correlate with those defined in the S-convention? In this section we address this question and in Section 3 we make use of that information to help us understand the lightcurve signatures of NPA rotators.

The NPA rotation is in either of two modes: the Short Axis Mode (SAM) where the short principal axis of the object circulates around the total rotational angular momentum vector (TRAMV) or the Long Axis Mode (LAM) where the long principal axis of the object circulates around the TRAMV (Julian 1987). In Appendix A of SA91, the properties of both LAMs and SAMs are described in terms of the L-convention whereas in Appendix A of this paper, we do the same in terms of the S-convention. We make use of these two Appendices to understand how L-convention component periods are correlated with those in the S-convention. In the following discussion, we adopt the subscript L to denote the L-convention while the subscript S denotes the S-convention. When neither L nor S is used in the subscript, it signifies that the relevant parameter corresponds to both L and S.

---

[2] The manifestation of component rotational periods in the lightcurve of an active comet which is in an NPA rotational state is different to that of a bare cometary nucleus and is much more complicated (e.g., depending on where the source regions for activity are located on the nucleus). Therefore, lightcurves of active comets are not considered in this paper.

## 2.1 $P_\psi$ and $P_\theta$ in the L-convention and the S-convention

### 2.1.1 In the L-convention

For LAMs, in the L-convention, the angles $\psi_L$ and $\theta_L$ are given by equations (A36) and (A37) respectively of SA91. They are

$$\psi_L = \mathrm{atan}2\left(\sqrt{\frac{I_i}{(I_i - I_l)}}\, \mathrm{sn}\,\tau,\, \sqrt{\frac{I_s}{(I_s - I_l)}}\, \mathrm{cn}\,\tau\right), \tag{1}$$

and

$$\theta_L = \cos^{-1}\left(\mathrm{dn}\,\tau \sqrt{\frac{I_l\left(I_s - \frac{M^2}{2E}\right)}{\frac{M^2}{2E}(I_s - I_l)}}\right) \tag{2}$$

where sn $\tau$, cn $\tau$, and dn $\tau$ are Jacobian elliptic functions and are periodic. FORTRAN function atan2($y$, $x$) represents $\tan^{-1}(y/x)$ with the associated signs for $y$ and $x$. $I_l$, $I_i$, and $I_s$ are the moment of inertia around the long, intermediate, and short principal axes. $M$ is the magnitude of the TRAMV and $E$ is the total kinetic energy of the NPA rotation.

The periods of sn $\tau$ and cn $\tau$ are the same and the period of dn $\tau$ is half of that (e.g., Abramowitz and Stegun 1964).

Therefore, for LAMs in the L-convention, we have

$$P_{\psi_L} = 2 P_{\theta_L}. \tag{3}$$

For SAMs in the L-convention, the angles $\psi_L$ and $\theta_L$ are given by equations (A60) and (A61) respectively of SA91, and they are

$$\psi_L = \mathrm{atan}2\left(\sqrt{\frac{I_i\left(I_s - \frac{M^2}{2E}\right)}{(I_s - I_i)}}\, \mathrm{sn}\,\tau,\, \sqrt{\frac{I_s\left(\frac{M^2}{2E} - I_l\right)}{(I_s - I_l)}}\, \mathrm{dn}\,\tau\right), \tag{4}$$

and

$$\theta_L = \cos^{-1}\left(\operatorname{cn}\tau \sqrt{\frac{I_l\left(I_s - \dfrac{M^2}{2E}\right)}{\dfrac{M^2}{2E}(I_s - I_l)}}\right). \tag{5}$$

Therefore, for SAMs in the L-convention,

$$P_{\psi_L} = P_{\theta_L}. \tag{6}$$

### 2.1.2 In the S-convention

For LAMs in the S-convention, the angles $\psi_S$ and $\theta_S$ are given by equations (A36) and (A37) respectively of Appendix A. Therefore, we obtain

$$P_{\psi_S} = P_{\theta_S}. \tag{7}$$

For SAMs in the S-convention, the corresponding angles for $\psi_S$ and $\theta_S$ are given by equations (A66) and (A67) respectively of Appendix A and we derive

$$P_{\psi_S} = 2P_{\theta_S}. \tag{8}$$

## 2.2 $P_\psi$ in the L-convention and the S-convention

A comparison of equation (1) above and equation (A36) of Appendix A shows that for LAMs, the period for angle $\psi$ is independent of whether it is defined in the L-convention or the S-convention (also see equation (A48) of Appendix A).

Similarly, by comparing equations (4) above and (A66) of Appendix A, for SAMs too it can be shown that the period for angle $\psi$ is independent of whether it is defined in the L-convention or the S-convention (also see equation (A74) of Appendix A).

Therefore, for all NPA rotational states, we have

$$P_{\psi_S} = P_{\psi_L}. \tag{9}$$

## 2.3 $P_\phi$ in the L-convention and the S-convention

From an examination of Appendix A of SA91 and Appendix A of this paper, it is clear that unlike the periods for angles $\psi$ and $\theta$, the periods for angle $\phi$ are derived based on the rate of change of $\phi$. Therefore, a comparison between the periods for angle $\phi$ in the L-convention and S-convention is not as straightforward as for angles $\psi$ and $\theta$.

However, during our study of periodicity signatures present in numerically generated lightcurves of NPA rotators, we found the following relationship, which we also confirmed numerically.

$$\frac{1}{P_{\phi_S}} = \frac{1}{P_{\phi_L}} + \frac{1}{P_{\psi_L}} \ . \tag{10}$$

Or, alternatively for the time-averaged rates, we have

$$\overline{\dot{\phi}_S} = \overline{\dot{\phi}_L} + \overline{\dot{\psi}_L} \ . \tag{11}$$

I.e., $\overline{\dot{\phi}_S} > \overline{\dot{\phi}_L}$.

From equation (10) it can be seen that $P_{\phi_S} < P_{\phi_L}$ and this will help one identify potential periodicity signatures in an actual lightcurve of an asteroid or a bare comet nucleus.

By examining the values for limiting cases (i.e., those corresponding to principal axis rotations) in the Appendix A of SA91 and Appendix A of this paper, it can be seen that they are consistent with equation (10).

## 2.4 Behavior of the ratio ($P_\psi / P_\phi$)

From equations (9) and (10), we find

$$\frac{P_{\psi_S}}{P_{\phi_S}} = \frac{P_{\psi_L}}{P_{\phi_L}} + 1 \ . \tag{12}$$

Again the limiting cases for this ratio (i.e., those corresponding to principal axis rotations around the short principal axis or the long principal axis) are consistent with Appendix A of SA91 and Appendix A of this paper (i.e., equation (A53) of SA91 and equation (A58) of this paper for LAMs; equation (A79) of SA91 and equation (A84) of this paper for SAMs).

# 3. Interpreting Periodicity Signatures Present in Simulated Lightcurves

In this section, we will investigate the periodicity signatures present in simulated lightcurves with component rotational periods known *a priori*. The periodicity signatures will be analyzed in both the L-convention as well as the S-convention. This will help us understand why certain periodicities show up in lightcurves but not others. We like to stress that this study in not intended to cover the entire parameter space for NPA rotators but to understand the periodicity signatures in terms of both the L-convention as well as the S-convention. An analysis covering the wider parameter space will be addressed in a separate paper (Mueller and Samarasinha, in preparation).

To illustrate the signatures of the NPA rotators in the L-convention and S-convention, we numerically generated ideal lightcurves with known L-convention component periods for different ellipsoidal shapes and various geometries. Ideal lightcurves are continuous (i.e. they have no gaps in time) and do not contain noise in their magnitudes. These lightcurves are then analyzed with a Fourier Transform (FT) to assess what kinds of signatures are present in the power spectrum. Figures 1 and 2 show power spectra for four different cases each for LAMs and for SAMs. The small body shape, component periods, and observing geometry for these cases are listed in Table 1. We used two different ellipsoidal shape models; shape 1 is closer to an oblate with semi-axes 6×5×3 km and shape 2 is closer to a prolate with semi-axes 6×3.5×3 km. We used four different geometries; for geo 1, the TRAMV is nearly parallel to (i.e., within a few degrees of) the line of sight, for geo 2, the geometry is almost perpendicular to the line of sight (i.e., within a few degrees of the perpendicular), and for geo 3 and geo 4, the geometries were chosen arbitrarily. The solar phase angle for all the lightcurves shown is ~2$^\circ$. We used the Hapke parameters for C-type asteroids (Helfenstein and Veverka 1989) for the scattering law for all numerically simulated lightcurves. In general, the use of a different photometric function will change the magnitude of the minima and maxima but not their locations in the rotational phase space and therefore will not significantly change the periodicity signatures in the FT for an ideal lightcurve. To further confirm this, we generated lightcurves (not shown) for the same conditions as earlier but assuming that the scattering is due to Lambertian particles, and as stated, there was no significant difference in the resulting periodicity signatures in the FT. Any difference is primarily limited to minor changes in the relative strengths of the periodicity signatures.

Table 1. Rotational and geometric parameters for SAM and LAM models in the L-system

| rotational state | case | shape | component periods [day] | geometry |
| --- | --- | --- | --- | --- |
| LAM | case 1 | shape 1 | $P_{\phi_L}=0.171$, $P_{\psi_L}=0.493$ | geo 2 |
| LAM | case 2 | shape 1 | $P_{\phi_L}=0.171$, $P_{\psi_L}=0.493$ | geo 4 |
| LAM | case 3 | shape 2 | $P_{\phi_L}=0.317$, $P_{\psi_L}=0.377$ | geo 2 |
| LAM | case 4 | shape 2 | $P_{\phi_L}=0.317$, $P_{\psi_L}=0.377$ | geo 3 |
| SAM | case 1 | shape 1 | $P_{\phi_L}=0.317$, $P_{\psi_L}=0.692$ | geo 1 |
| SAM | case 2 | shape 1 | $P_{\phi_L}=0.317$, $P_{\psi_L}=0.692$ | geo 2 |
| SAM | case 3 | shape 2 | $P_{\phi_L}=0.171$, $P_{\psi_L}=0.613$ | geo 3 |
| SAM | case 4 | shape 2 | $P_{\phi_L}=0.171$, $P_{\psi_L}=0.613$ | geo 4 |

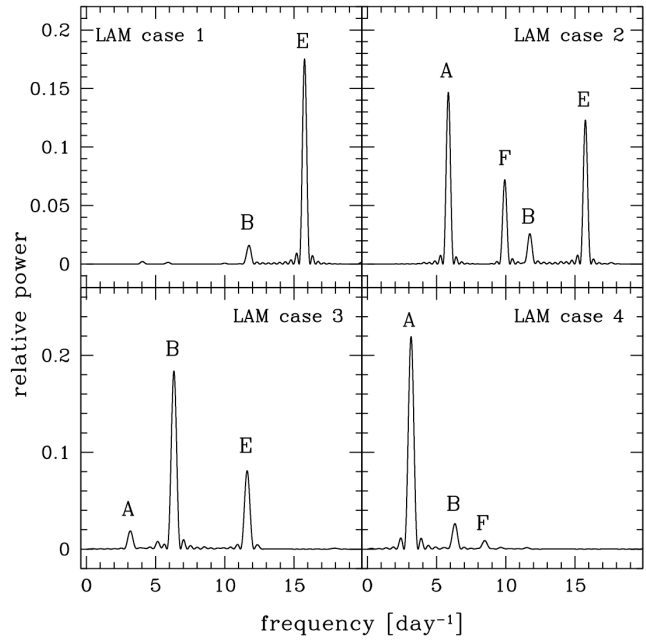

Figure 1: Periodicity signatures found in the ideal lightcurves for the four LAM cases described in Table 1.

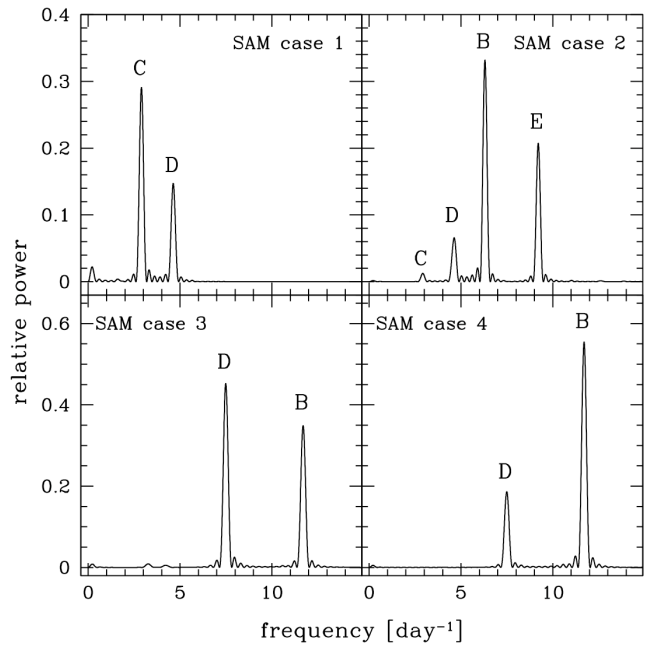

Figure 2: Periodicity signatures found in the ideal lightcurves for the four SAM cases described in Table 1.

Figures 1 and 2 clearly show that there are multiple periodicities present in the ideal lightcurves of both LAM and SAM rotational states after the application of a FT. In Table 2 these periodicity signatures are listed in terms of the known L-convention periods as well as the corresponding S-convention values using equations (9) and (10).

Table 2. Identification of rotational signatures shown in Figures 1 and 2.

| peak | rotational signature | LAM cases | SAM cases |
|------|----------------------|-----------|-----------|
| A | $1/P_{\phi_L} = 1/P_{\phi_S} - 1/P_{\psi_S}$ | 2, 3, 4 | ⋯ |
| B | $2/P_{\phi_L} = 2/P_{\phi_S} - 2/P_{\psi_S}$ | 1, 2, 3, 4 | 2, 3, 4 |
| C | $2/P_{\psi_L} = 2/P_{\psi_S}$ | ⋯ | 1, 2 |
| D | $1/P_{\phi_L} + 1/P_{\psi_L} = 1/P_{\phi_S}$ | ⋯ | 1, 2, 3, 4 |
| E | $2/P_{\phi_L} + 2/P_{\psi_L} = 2/P_{\phi_S}$ | 1, 2, 3 | 2 |
| F | $1/P_{\phi_L} + 2/P_{\psi_L} = 1/P_{\phi_S} + 1/P_{\psi_S}$ | 2, 4 | ⋯ |

It can be seen from Table 2 that four prominent and frequently found periodicities in these lightcurves are $1/P_{\phi_L}$, $2/P_{\phi_L}$, $1/P_{\phi_S}$, and $2/P_{\phi_S}$ although not all these signatures are present in a given single lightcurve. Some of these signatures would have been identified only as combinations of periods in the L-convention if we were not aware of equation (10) (i.e., $(1/P_{\phi_L} + 1/P_{\psi_L})$ instead of $1/P_{\phi_S}$ and $(2/P_{\phi_L} + 2/P_{\psi_L})$ instead of $2/P_{\phi_S}$). The fact that some signatures containing both $1/P_\phi$ and $1/P_\psi$ can indeed be defined in terms of a single component period in a different convention is helpful when one tries to interpret periodicity signatures in actual lightcurves (e.g., Pravec et al. 2005, 2014). However, additional signatures corresponding to combinations containing both $1/P_\phi$ and $1/P_\psi$ that cannot be identified with a single period in a different convention are still viable (e.g., such as the signature corresponding to the peak F above). In addition to these, periodicity signatures corresponding to the harmonics of $1/P_\psi$ are also possible. It is pertinent to point out that the lightcurve signatures shown in Kaasalainen (2001) as well as the additional numerical simulations by us not shown here (e.g., Mueller et al. 2002, Mueller and Samarasinha 2012) are consistent with these results.

We caution the reader that the list of periodicities listed in Table 2 does not necessarily cover all the possible periodicities as we only show results for a limited representative sample of lightcurves. In addition, we find that lightcurves for irregular shapes, which are more representative of the shapes of actual asteroids or cometary nuclei, are capable of generating additional periodicity signatures that a triaxial ellipsoidal shape might not reveal due to its axial symmetry.

Next we like to point out that the object shape and the observing geometry have notable effects on which periodicities may get manifested in the lightcurves as well as on the relative strengths of the signatures. For example, LAM case 1 and LAM case 3 in Table 1 have different object shapes but the same observing geometry where the TRAMV is nearly perpendicular to the line of sight. According to Figure 1, both cases reveal nearly the same periodicity signatures but of different strengths despite being in two different LAM rotational states. On the other hand, in Table 1, we identify that the following pairs of cases are having the same object shape but different observing geometries: (LAM case 1 and LAM case 2), (LAM case 3 and LAM case 4), (SAM case 1 and SAM case 2), and (SAM case 3 and SAM case 4). Except for the last pair where the two geometries are arbitrary, the signatures present in the lightcurves are extremely sensitive to the observing geometry.

As ideal lightcurves are rarely observed from the ground, we approximated observing conditions by using data only for 8 hours per day, randomly dropping about 30% of the remaining data, and adding random noise with a Gaussian distribution with a standard deviation σ of 0.05 mag in the magnitude. The resulting simulated observational data points are shown in Figures 3 and 4. We also show the underlying ideal lightcurves in the figures for the reader to visualize the effects of random noise and random sampling during a night as well as the diurnal gaps.

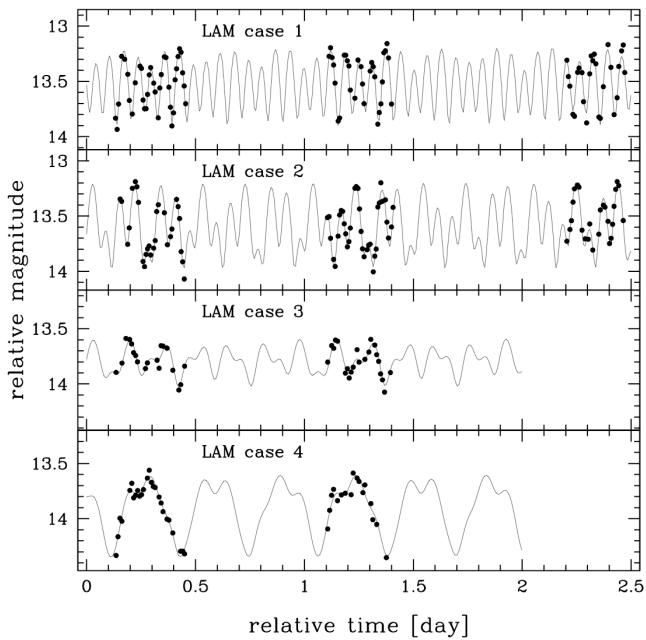

Figure 3: The lightcurves corresponding to the LAM cases where we simulate more realistic observational conditions by considering observational gaps and noise. The simulated data points are shown as solid circles. The solid lines show the corresponding ideal lightcurves.

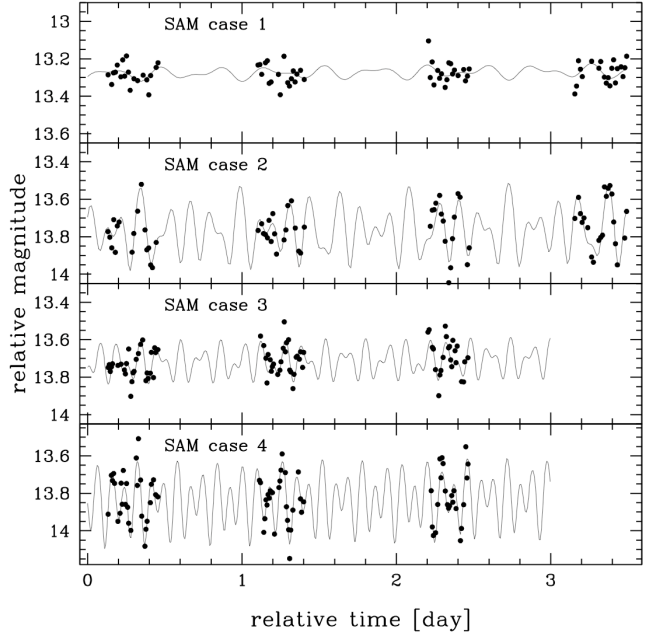

Figure 4: The lightcurves corresponding to the SAM cases where we simulate more realistic observational conditions by considering observational gaps and noise. The simulated data points are shown as solid circles. The solid lines show the corresponding ideal lightcurves.

The corresponding FTs are shown on the top in Figures 5 and 6. As these lightcurves have substantial diurnal gaps, the FTs show daily aliasing as well as harmonics of the component periods. We call these FT power spectra "dirty spectra". To eliminate these artifacts, we apply a clean algorithm (Roberts et al. 1987) with a window function (Belton and Gandhi 1988). The resulting power spectra are called "clean spectra" and are shown on the bottom in Figures 5 and 6. These figures illustrate that in all cases except for SAM case 1, essentially the same signatures are present as in the FTs of the ideal lightcurves shown in Figures 1 and 2. However, sometimes certain signatures may have much smaller relative power or can be absent. An asterisk next to the identifying letter signifies that the corresponding signature seen for the ideal lightcurve is not present. All the other signatures (except for SAM case 1) that are present in Figures 5 and 6 but not identified explicitly are daily aliases that have not been properly cleaned out. The dirty and clean spectra for SAM case 1 show many signatures not related to the component periods. Since the corresponding lightcurve (Figure 4, top) has a relatively small amplitude, yet the photometric errors are comparable to other high-amplitude lightcurves, it is harder to recover only the underlying periodic signatures in this case. Also, in some cases, the weaker signatures may disappear or may be only marginally detectable (e.g., LAM cases 1 or 4) leading one to infer principal-axis rotation. We like to stress that Figures 5 and 6 are only an illustration to demonstrate that the periodicity signatures identified in Table 2 are still mostly present in the non-ideal lightcurves as well. Yet, which of these signatures are present in what relative power is dependent on the object shape, rotational state, observing geometry, temporal sampling, and the noise levels.

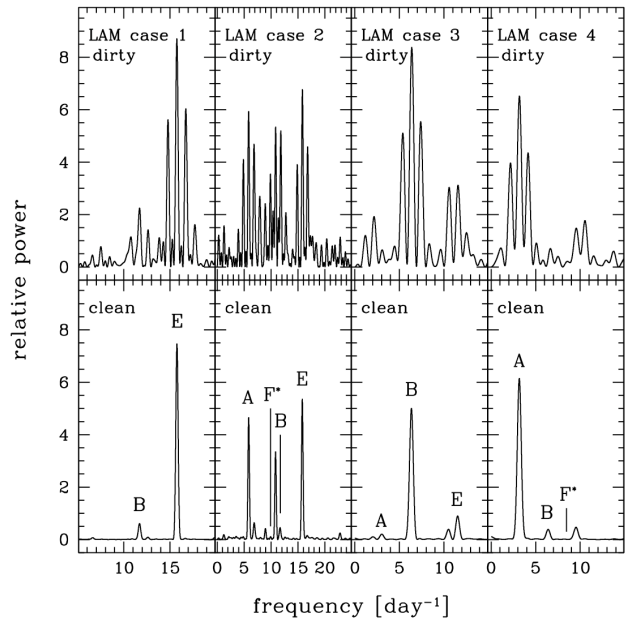

Figure 5: The corresponding Fourier Transforms for the LAM cases for the simulated lightcurves with observational gaps and noise shown in figure 3. Top: FT for dirty spectra; Bottom: FT for clean spectra.

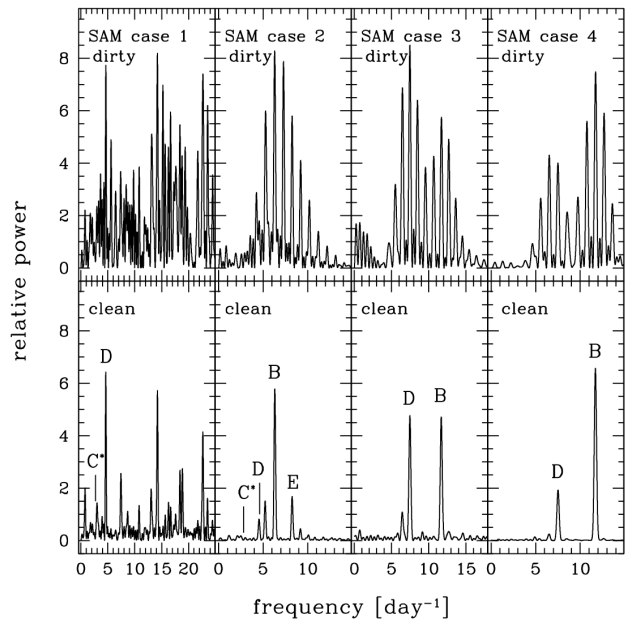

Figure 6: The corresponding Fourier Transforms for the SAM cases for the simulated lightcurves with observational gaps and noise shown in figure 4.

# 4. Conclusions

The main conclusions of this paper are the following.

(a) We show that the component periods of NPA rotation in the L-convention and the S-convention are correlated with each other, thus enabling us to better interpret periodicity signatures present in the lightcurves of NPA rotators.

   (1) The periods $P_\psi$ and $P_\theta$ are related by either $P_\psi = P_\theta$ or $P_\psi = 2P_\theta$ depending on whether it is a LAM or a SAM or is expressed in the L-convention or the S-convention.

   (2) The periods for the Euler angle $\psi$ in the L-convention and the S-convention are related by $P_{\psi_S} = P_{\psi_L}$.

   (3) The periods for the Euler angle $\phi$ in the L-convention and the S-convention are related by $\dfrac{1}{P_{\phi_S}} = \dfrac{1}{P_{\phi_L}} + \dfrac{1}{P_{\psi_L}}$.

Therefore, when identifying the lightcurve signatures, we advocate that researchers should always specify the convention in which component periods are defined.

(b) In the lightcurves of NPA rotators, the periodicity signatures for $1/P_{\phi_L}, 2/P_{\phi_L}, 1/P_{\phi_S}$, and $2/P_{\phi_S}$ are present frequently and prominently but not in each and every lightcurve. In addition, periodicity signatures corresponding to the harmonics of $1/P_\psi$ and the combinations containing both $1/P_\phi$ and $1/P_\psi$ are also possible.

(c) The periodicity signatures present in the lightcurves are sensitive to the observing geometry, object shape, particular NPA rotational state, and the observing circumstances such as temporal sampling and noise level. In certain cases (including sparsely sampled or noisy lightcurves), it is possible that the observer will potentially misidentify an NPA rotator as a principal-axis rotator.

# Acknowledgements

We acknowledge support from NASA's Outer Planets Research Program through grant number NNX10AP95G. We thank Ruvini Samarasinha for drafting Figure A1. We thank the referees Drs. Pedro Gutiérrez and Petr Pravec for their thoughtful comments that helped improve the quality of the paper. This is PSI Contribution Number 624.

# Appendix A

In this Appendix, we provide the analytical expressions for the non-principal-axis (NPA) rotation from the perspective of the S-convention that is suited to describe the component rotations associated with the motion of the short principal axis. Such a convention is widely used in textbooks that discuss dynamics of rigid bodies (e.g., Landau and Lifshitz 1976) and is especially suitable when describing the rotational dynamics of the Earth. Also, it is used by Kaasalainen (2001) to describe the NPA rotation of asteroids. In this Appendix, we will follow an approach analogous to that in the Appendix of Samarasinha and A'Hearn (1991) (SA91) in which the NPA component rotations are described in terms of the motion of the long principal axis (in the L-convention). In SA91, we have chosen such a convention, as that is appropriate to interpret the rotational motion of an elongated object such as the nucleus of comet 1P/Halley — the focus of that paper. We also attempt to present the derivations in a format that will facilitate direct comparisons of the equations in this Appendix with the equations in the Appendix of SA91. In this paper, we select the subscript $S$ (e.g., for symbols denoting the component periods) to denote the S-convention whereas the subscript $L$ is used to denote the parameters associated with the L-convention used in SA91. It is pertinent to mention here that the S-convention is simply the widely used *zxz* convention (also known as the *x* convention; e.g., Goldstein 1980) in standard mechanics textbooks in the context of defining Euler angles.

The subscripts *l, i,* and *s* denote the long, intermediate, and short principal axes respectively and *lis* represents a right-handed body frame coordinate system with its origin at the center of mass similar to SA91. The *XYZ* system represents the external reference frame of a co-orbiting observer also centered at the center of mass of the body (Figure A1). As in SA91, the total rotational angular momentum vector (TRAMV), **M**, which is fixed in the external reference frame when no external torques are present, is aligned with the *Z*-axis. The Euler angles $\phi_S$, $\theta_S$, and $\psi_S$ describe the rotational motion of the body with respect to the external observer.

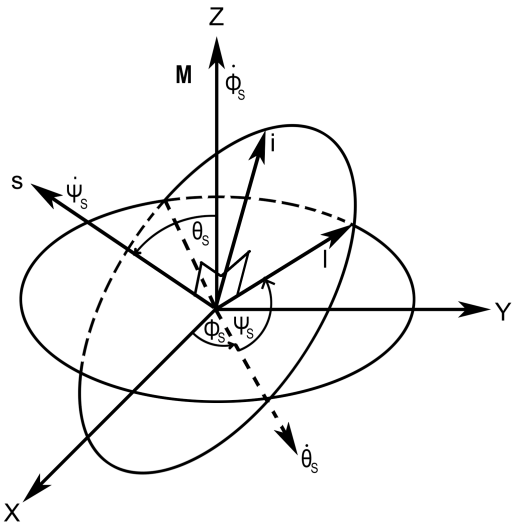

Figure A1: The body frame coordinate system (*lis* right-handed system) and the *XYZ* inertial frame coordinate system when describing the rotation in terms of the S-convention. The *Z*-axis is aligned with the TRAMV, **M.** The angles $\phi_S$, $\theta_S$, and $\psi_S$ are Euler angles and the time derivatives of them are used to define the periods associated with the component motions of the NPA rotation.

The rate of change of the TRAMV, **M**, in the external frame is given by

$$\left(\frac{d\mathbf{M}}{dt}\right)_{ext} = 0. \tag{A1}$$

The rate of change of **M** in the body frame is given by

$$\left(\frac{d\mathbf{M}}{dt}\right)_{body} + \Omega \times \mathbf{M} = \left(\frac{d\mathbf{M}}{dt}\right)_{ext} = 0, \tag{A2}$$

where $\Omega$ is the instantaneous angular velocity vector of the body frame with respect to the external frame; $\Omega$ is not constant in direction nor magnitude and moves around **M** at a variable rate in the external frame. If $l$, $i$, and $s$ components of **M** are denoted by $M_k$ and are given by

$$M_k = I_k \Omega_k \tag{A3}$$

then,

$$\frac{dM_k}{dt} = I_k \frac{d\Omega_k}{dt}. \tag{A4}$$

From equations (A2) and (A4),

$$\frac{d\Omega_l}{dt} = \frac{\Omega_i \Omega_s}{I_l}(I_i - I_s), \tag{A5}$$

$$\frac{d\Omega_i}{dt} = \frac{\Omega_s \Omega_l}{I_i}(I_s - I_l), \tag{A6}$$

and

$$\frac{d\Omega_s}{dt} = \frac{\Omega_l \Omega_i}{I_s}(I_l - I_i). \tag{A7}$$

From conservation of energy,

$$I_l \Omega_l^2 + I_i \Omega_i^2 + I_s \Omega_s^2 = 2E, \tag{A8}$$

where $E$ is the total rotational kinetic energy of the body.

From conservation of angular momentum,

$$I_l^2 \Omega_l^2 + I_i^2 \Omega_i^2 + I_s^2 \Omega_s^2 = M^2. \tag{A9}$$

Using equations (A8) and (A9),

$$\Omega_l^2 = \frac{\left[2E\left(I_s - \frac{M^2}{2E}\right) - I_i(I_s - I_i)\Omega_i^2\right]}{I_l(I_s - I_l)} \quad (A10)$$

and

$$\Omega_s^2 = \frac{\left[2E\left(\frac{M^2}{2E} - I_l\right) - I_i(I_i - I_l)\Omega_i^2\right]}{I_s(I_s - I_l)} . \quad (A11)$$

From equations (A6), (A10), and (A11),

$$\frac{d\Omega_i}{dt} = \frac{\sqrt{\left[2E\left(I_s - \frac{M^2}{2E}\right) - I_i(I_s - I_i)\Omega_i^2\right] \times \left[2E\left(\frac{M^2}{2E} - I_l\right) - I_i(I_i - I_l)\Omega_i^2\right]}}{I_i\sqrt{I_l I_s}} \quad (A12)$$

### *Behavior of Euler Angles and the Parameter Space*

The Euler angles $\phi_S$, $\theta_S$, and $\psi_S$ can be expressed in terms of the component angular velocities in the body frame $\Omega_l$, $\Omega_i$, and $\Omega_s$ as follows.

For Euler angles $\theta_S$ and $\psi_S$ we have

$$M \sin\theta_S \sin\psi_S = M_l = I_l \Omega_l \quad (A13)$$

and

$$M \sin\theta_S \cos\psi_S = M_i = I_i \Omega_i . \quad (A14)$$

Since $0 \le \theta_S \le \pi$, $\sin \theta_S \ge 0$, and $M > 0$, for example, the FORTRAN function *atan2* (as well as the corresponding function in many other computer languages and math packages) will uniquely determine $\psi_S$ from

$$\psi_S = \text{atan2}\ (I_l\Omega_l,\ I_i\Omega_i) . \quad (A15)$$

The angle $\theta_S$ can be calculated from

$$M \cos\theta_S = M_s = I_s \Omega_s .$$

i.e.,

$$\theta_S = \cos^{-1}\left(\frac{I_s \Omega_s}{M}\right). \tag{A16}$$

The other Euler angle, $\phi_S$, and its time derivative $\dot{\phi}_S$ can be determined from

$$\Omega_l = \dot{\phi}_S \sin\theta_S \sin\psi_S + \dot{\theta}_S \cos\psi_S \tag{A17}$$

and

$$\Omega_i = \dot{\phi}_S \sin\theta_S \cos\psi_S - \dot{\theta}_S \sin\psi_S. \tag{A18}$$

Solving for $\dot{\phi}_S$ results in

$$\dot{\phi}_S = \frac{\Omega_l \sin\psi_S + \Omega_i \cos\psi_S}{\sin\theta_S}. \tag{A19}$$

Using equations (A9), (A13), (A14), and (A16), equation (A19) can be rewritten as

$$\dot{\phi}_S = M\left[\frac{I_l \Omega_l^2 + I_i \Omega_i^2}{I_l^2 \Omega_l^2 + I_i^2 \Omega_i^2}\right]. \tag{A20}$$

So, $\phi_S = \int \dot{\phi}_S dt$. \hfill (A21)

The equations (A17) and (A18) can be used to derive the time derivative of the Euler angle $\theta_S$ and is given by

$$\dot{\theta}_S = \Omega_l \cos\psi_S - \Omega_i \sin\psi_S. \tag{A22}$$

The time derivative of the Euler angle $\psi_S$ can be obtained from

$$\Omega_s = \dot{\psi}_S + \dot{\phi}_S \cos\theta_S. \tag{A23}$$

From equations (A16), (A20), and (A23), we have

$$\dot{\psi}_S = -\Omega_s \left[\frac{(I_s - I_l)I_l \Omega_l^2 + (I_s - I_i)I_i \Omega_i^2}{I_l^2 \Omega_l^2 + I_i^2 \Omega_i^2}\right]. \tag{A24}$$

Note that since $I_s \geq I_i \geq I_l$, when $\Omega_s > 0$, $\dot{\psi}_S < 0$ (i.e., the sense of motions for Euler angles $\phi_S$ and $\psi_S$ would be opposite unlike in SA91).

Using equation (A3), equation (A8) can be rewritten as

$$\frac{M_l^2}{I_l} + \frac{M_i^2}{I_i} + \frac{M_s^2}{I_s} = 2E. \tag{A25}$$

Also, using equation (A3), equation (A9) can be rewritten as

$$M_l^2 + M_i^2 + M_s^2 = M^2. \tag{A26}$$

Since $I_l \leq I_i \leq I_s$, we have $I_l \leq M^2/2E \leq I_s$ and when $I_l \leq M^2/2E < I_i$, the motion is a Long Axis Mode (LAM) and when $I_i < M^2/2E \leq I_s$, the motion is a Short Axis Mode (SAM).

For an equivalent triaxial ellipsoid of axial lengths $L_l$, $L_i$, and $L_s$ and mass $\mu$,

$$I_l = \frac{\mu}{20}(L_i^2 + L_s^2), \tag{A27}$$

$$I_i = \frac{\mu}{20}(L_l^2 + L_s^2), \tag{A28}$$

and

$$I_s = \frac{\mu}{20}(L_l^2 + L_i^2). \tag{A29}$$

As pointed out in SA91, the motion can be described in terms of $I_l/\mu$, $I_i/\mu$, $I_s/\mu$, $E/\mu$, and $M/\mu$. I.e., the motion can be expressed in terms of $L_l$, $L_i$, $L_s$ and the two free parameters $E/\mu$ and $M^2/(2E\mu)$.

*Long Axis Modes (LAMs)*

For LAMs,
$$I_l \leq \frac{M^2}{2E} < I_i. \tag{A30}$$
Define an independent variable $\tau$ of time t, and a constant of the motion, $k^2$ ($0 \leq k^2 < 1$), by

$$\tau = t\sqrt{\frac{2E(I_i - I_l)\left(I_s - \dfrac{M^2}{2E}\right)}{I_l I_i I_s}} \tag{A31}$$

and

$$k^2 = \frac{(I_s - I_i)\left(\dfrac{M^2}{2E} - I_l\right)}{(I_i - I_l)\left(I_s - \dfrac{M^2}{2E}\right)}.$$ (A32)

Note that both $\tau$ and $k^2$ are defined identical to how they are defined in SA91. When $M^2/2E \to I_l$, $k^2 \to 0$ while when $M^2/2E \to I_i$, $k^2 \to 1$.

Then equations (A10), (A11), and (12) can be rewritten to express $\Omega_l$, $\Omega_i$, and $\Omega_s$ in terms of the Jacobian elliptic functions sn $\tau$, cn $\tau$, and dn $\tau$.

$$\Omega_l = \mathrm{dn}\,\tau \sqrt{\frac{2E\left(I_s - \dfrac{M^2}{2E}\right)}{I_l(I_s - I_l)}},$$ (A33)

$$\Omega_i = \mathrm{sn}\,\tau \sqrt{\frac{2E\left(\dfrac{M^2}{2E} - I_l\right)}{I_i(I_i - I_l)}},$$ (A34)

and

$$\Omega_s = \mathrm{cn}\,\tau \sqrt{\frac{2E\left(\dfrac{M^2}{2E} - I_l\right)}{I_s(I_s - I_l)}}.$$ (A35)

Again, equations (A33), (A34), and (A35) are identical to those in SA91 and this is consistent with the expectation that component angular velocities in the body frame should be independent of the particular convention chosen.

Since dn $\tau > 0$ at all times, based on equation (A33), $\Omega_l > 0$ at all times for LAMs whereas that is not the case for $\Omega_i$ and $\Omega_s$.

Then equations (A15), (A16), and (A20) can be rewritten as

$$\psi_S = \mathrm{atan2}\left(\sqrt{\frac{I_l\left(I_s - \dfrac{M^2}{2E}\right)}{(I_s - I_l)}}\,\mathrm{dn}\,\tau,\, \sqrt{\frac{I_i\left(\dfrac{M^2}{2E} - I_l\right)}{(I_i - I_l)}}\,\mathrm{sn}\,\tau\right),$$ (A36)

$$\theta_S = \cos^{-1}\left(\operatorname{cn}\tau \sqrt{\frac{I_s\left(\frac{M^2}{2E} - I_l\right)}{\frac{M^2}{2E}(I_s - I_l)}}\right), \qquad (A37)$$

and

$$\dot{\phi}_S = M\left[\frac{\left(I_s - \frac{M^2}{2E}\right) + \left(\frac{M^2}{2E} - I_l\right)\operatorname{sn}^2\tau}{I_l\left(I_s - \frac{M^2}{2E}\right) + I_s\left(\frac{M^2}{2E} - I_l\right)\operatorname{sn}^2\tau}\right]. \qquad (A38)$$

From equation (A36), we see that the angle $\psi_S$ oscillates between $(\psi_S)_{\min}$ (when sn $\tau$ =1) and $(\psi_S)_{\max}$ (when sn $\tau$ = -1) on either side of $\psi_S = \pi/2$ where

$$(\psi_S)_{\min} = \tan^{-1}\left(\sqrt{\frac{I_l\left(I_i - \frac{M^2}{2E}\right)}{I_i\left(\frac{M^2}{2E} - I_l\right)}}\right) \qquad (A39)$$

and

$$(\psi_S)_{\max} = \tan^{-1}\left(-\sqrt{\frac{I_l\left(I_i - \frac{M^2}{2E}\right)}{I_i\left(\frac{M^2}{2E} - I_l\right)}}\right). \qquad (A40)$$

Therefore, the amplitude of this oscillatory motion of $\psi_S$ is given by

$$A_{\psi_S} = \tan^{-1}\left(\sqrt{\frac{I_i\left(\frac{M^2}{2E} - I_l\right)}{I_l\left(I_i - \frac{M^2}{2E}\right)}}\right). \qquad (A41)$$

Note that when $M^2/2E \to I_l$, the amplitude of the oscillatory motion of $\psi_S$ approaches zero. I.e., $\psi_S \to \pi/2$. On the other hand, when $M^2/2E \to I_i$, the amplitude of the oscillatory motion of $\psi_S$ approaches $\pi/2$.

From equation (A37), we can see that $\theta_S$ oscillates between $(\theta_S)_{min}$ (when cn $\tau =1$) and $(\theta_S)_{max}$ (when cn $\tau = -1$) on either side of $\theta_S = \pi/2$ where

$$(\theta_S)_{min} = \cos^{-1}\left(\sqrt{\frac{I_s\left(\frac{M^2}{2E} - I_l\right)}{\frac{M^2}{2E}(I_s - I_l)}}\right), \tag{A42}$$

and

$$(\theta_S)_{max} = \cos^{-1}\left(-\sqrt{\frac{I_s\left(\frac{M^2}{2E} - I_l\right)}{\frac{M^2}{2E}(I_s - I_l)}}\right). \tag{A43}$$

The amplitude of this oscillatory motion of $\theta_S$ is given by

$$A_{\theta_S} = \sin^{-1}\left(\sqrt{\frac{I_s\left(\frac{M^2}{2E} - I_l\right)}{\frac{M^2}{2E}(I_s - I_l)}}\right). \tag{A44}$$

Note that when $M^2/2E \to I_l$, the amplitude of the oscillatory motion of $\theta_S$ approaches zero. I.e., $\theta_S = \pi/2$. However, when $M^2/2E \to I_i$, the amplitude of the oscillatory motion of $\theta_S$ does not approach $\pi/2$ but a value that is less than $\pi/2$ (except when $I_s=I_i$).

From equation (A38), we can see that $\dot{\phi}_S$ oscillates between $(\dot{\phi}_S)_{min}$ (when sn $\tau = \pm 1$) and $(\dot{\phi}_S)_{max}$ (when sn $\tau = 0$) with

$$(\dot{\phi}_S)_{min} = \frac{M}{\left(\frac{M^2}{2E}\right)} \tag{A45}$$

and

$$(\dot{\phi}_S)_{max} = \frac{M}{I_l}. \tag{A46}$$

Note that when $M^2/2E \to I_l$, $\dot{\phi}_S$ approaches the constant rate of $M/I_l$.

For LAMs in the S-convention, $\phi_S$ is the only angle that circulates through a full $2\pi$ radians (analogous to $\phi_L$ for SAMs in the L-convention).

The period of angle $\psi_S$, $P_{\psi_S}$, is given by (cf. Landau and Lifshitz 1976, pp. 118-119)

$$P_{\psi_S} = 4\sqrt{\frac{I_l I_i I_s}{2E(I_i - I_l)\left(I_s - \frac{M^2}{2E}\right)}} \int_0^{\pi/2} \frac{du}{\sqrt{1 - k^2 \sin^2 u}} \quad (A47)$$

and is the same as the period of sn $\tau$.

Note this period for angle $\psi_S$ in the S-convention is the same as that of the angle $\psi_L$ in the L–convention for LAMs (compare equation (A47) above with equation (A45) of SA91). Therefore, for LAMs,

$$P_{\psi_S} = P_{\psi_L}. \quad (A48)$$

The period $P_{\psi_S}$ will have its minimum and maximum values when $k^2 = 0$ and $k^2 \to 1$, respectively.

When $M^2/2E = I_l$, $k^2 = 0$ and

$$P_{\psi_S} = 2\pi \sqrt{\frac{I_l I_i I_s}{2E(I_i - I_l)(I_s - I_l)}}.$$

i.e.,

$$(P_{\psi_S})_{\min} = \frac{2\pi I_l}{M} \sqrt{\frac{I_i I_s}{(I_i - I_l)(I_s - I_l)}}. \quad (A49)$$

When $M^2/2E \to I_i$, $k^2 \to 1$ and the integral on the right side of equation (A47) $\to \infty$.
i.e.,

$$(P_{\psi_S})_{\max} \to \infty. \quad (A50)$$

Therefore for LAMs,

$$\frac{2\pi I_l}{M} \sqrt{\frac{I_i I_s}{(I_i - I_l)(I_s - I_l)}} \leq P_{\psi_S} < \infty \quad (A51)$$

From equation (A38), it can be seen that $\dot{\phi}_S$ is periodic with period $P_\psi/2$. The angle $\phi_S$ is aperiodic unless $M^2/2E = I_l$ or $\int_0^{P_{\psi_S}/2} \dot{\phi}_S \, dt = 2\pi$ (this latter condition may not exist for certain values of axial lengths $L_l$, $L_i$, and $L_s$). One can define a time-averaged period for $\phi_S$, $P_{\overline{\phi_S}}$, where $P_{\bar{\phi}_S} = 2\pi/(\dot{\phi}_S)_{mean}$ and $(\dot{\phi}_S)_{mean} = (2/P_{\psi_S}) \int_0^{P_{\psi_S}/2} \dot{\phi}_S \, dt$. (Note: Hereafter, for simplicity, we will denote $P_{\bar{\phi}_S}$ simply by $P_{\phi_S}$). Based on the properties of sn $\tau$, from equation (A38) we can deduce that $P_{\phi_S}$ is minimum when $k^2 \to 0$ (i.e., when sn $\tau \to \sin \tau$ and $M^2/2E \to I_l$) and maximum when $k^2 \to 1$ (i.e., when sn $\tau \to \tanh \tau$ and $M^2/2E \to I_i$).

Therefore from equation (A38), we have

$$(P_{\phi_S})_{min} = \frac{2\pi I_l}{M} \tag{A52}$$

and

$$(P_{\phi_S})_{max} = \frac{2\pi I_i}{M}. \tag{A53}$$

Therefore for LAMs,

$$\frac{2\pi I_l}{M} \leq P_{\phi_S} \leq \frac{2\pi I_i}{M}. \tag{A54}$$

Note that in the L-convention (unlike $(P_{\phi_S})_{min}$ and $(P_{\phi_S})_{max}$ in the S-convention), for LAMs, $(P_{\phi_L})_{min}$ occurs when $M^2/2E \to I_i$ (i.e., $k^2 \to 1$) and $(P_{\phi_L})_{max}$ occurs when $M^2/2E \to I_l$ (i.e., when $k^2 \to 0$) (cf. SA91).

Still in the S-convention, $\left(\frac{P_{\psi_S}}{P_{\phi_S}}\right)_{min}$ occurs when $k^2 \to 0$ and $\left(\frac{P_{\psi_S}}{P_{\phi_S}}\right)_{max}$ occurs when $k^2 \to 1$.

From equations (A49) and (A52), we obtain

$$\left(\frac{P_{\psi_S}}{P_{\phi_S}}\right)_{min} = \sqrt{\frac{I_i I_s}{(I_i - I_l)(I_s - I_l)}} \tag{A55}$$

whereas from equations (A50) and (A53), we have

$$\left(\frac{P_{\psi_S}}{P_{\phi_S}}\right)_{max} \to \infty. \tag{A56}$$

i.e.,

$$\left(\frac{P_{\psi_S}}{P_{\phi_S}}\right) \geq \sqrt{\frac{I_i I_s}{(I_i - I_l)(I_s - I_l)}}. \tag{A57}$$

Using equations (A27), (A28), and (A29) we can rewrite equation (A57) as

$$\left(\frac{P_{\psi_s}}{P_{\phi_s}}\right) \geq \sqrt{\frac{(L_l^2 + L_s^2)(L_l^2 + L_i^2)}{(L_l^2 - L_s^2)(L_l^2 - L_i^2)}}. \tag{A58}$$

Since $\sqrt{\dfrac{(L_l^2 + L_s^2)(L_l^2 + L_i^2)}{(L_l^2 - L_s^2)(L_l^2 - L_i^2)}} > 1$,

$$\left(\frac{P_{\psi_s}}{P_{\phi_s}}\right) > 1. \tag{A59}$$

*Short Axis Modes (SAMs)*

For SAMs,
$$I_i < \frac{M^2}{2E} \leq I_s. \tag{A60}$$

Analogues to equations (A31) and (A32), $\tau$ and $k^2$ ($0 \leq k^2 < 1$) can be defined by

$$\tau = t\sqrt{\frac{2E(I_s - I_i)\left(\dfrac{M^2}{2E} - I_l\right)}{I_l I_i I_s}} \tag{A61}$$

and

$$k^2 = \frac{(I_i - I_l)\left(I_s - \dfrac{M^2}{2E}\right)}{(I_s - I_i)\left(\dfrac{M^2}{2E} - I_l\right)}. \tag{A62}$$

Again, note that both $\tau$ and $k^2$ are defined identical to how they are defined in SA91. Then equations (A10), (A11), and (A12) can be rewritten to express $\Omega_l$, $\Omega_I$, and $\Omega_s$ as

$$\Omega_l = \operatorname{cn} \tau \sqrt{\frac{2E\left(I_s - \dfrac{M^2}{2E}\right)}{I_l(I_s - I_l)}}, \tag{A63}$$

$$\Omega_i = \operatorname{sn} \tau \sqrt{\frac{2E\left(I_s - \frac{M^2}{2E}\right)}{I_i(I_s - I_i)}},$$
(A64)

and

$$\Omega_s = \operatorname{dn} \tau \sqrt{\frac{2E\left(\frac{M^2}{2E} - I_l\right)}{I_s(I_s - I_l)}}.$$
(A65)

Again, equations (A63), (A64), and (A65) are identical to corresponding equations in SA91 and this is consistent with the expectation that component angular velocities in the body frame should be independent of the particular convention that one has chosen.

Since dn $\tau > 0$ at all times, based on equation (A65), $\Omega_s > 0$ at all times for SAMs; however, that is not the case for $\Omega_l$ and $\Omega_i$. Since $\Omega_s > 0$, based on equation (A24), for SAMs, $\dot{\psi}_S < 0$ at all times. I.e., The sense of motions for $\phi_S$ and $\psi_S$ are opposite to each other for SAMs.

The equations (A15), (A16), and (A20) can be rewritten as

$$\psi_S = \operatorname{atan2}\left(\sqrt{\frac{I_l\left(I_s - \frac{M^2}{2E}\right)}{(I_s - I_l)}} \operatorname{cn} \tau, \sqrt{\frac{I_i\left(I_s - \frac{M^2}{2E}\right)}{(I_s - I_i)}} \operatorname{sn} \tau\right),$$
(A66)

$$\theta_S = \cos^{-1}\left(\operatorname{dn} \tau \sqrt{\frac{I_s\left(\frac{M^2}{2E} - I_l\right)}{\frac{M^2}{2E}(I_s - I_l)}}\right),$$
(A67)

and

$$\dot{\phi}_S = M\left[\frac{(I_s - I_i) + (I_i - I_l)\operatorname{sn}^2 \tau}{I_l(I_s - I_i) + I_s(I_i - I_l)\operatorname{sn}^2 \tau}\right].$$
(A68)

From equations (A66) and (A68) we can see that both angles $\psi_S$ and $\phi_S$ circulate through full $2\pi$ radians albeit in opposite senses.

From equation (A67), we can see that $\theta_S$ oscillates between $(\theta_S)_{min}$ (when $\operatorname{dn}\tau = 1$) and $(\theta_S)_{max}$ (when $\operatorname{dn}\tau = \sqrt{1-k^2}$).

Therefore, we have

$$(\theta_S)_{min} = \cos^{-1}\left(\sqrt{\frac{I_s\left(\frac{M^2}{2E}-I_l\right)}{\frac{M^2}{2E}(I_s-I_l)}}\right), \quad (A69)$$

and

$$(\theta_S)_{max} = \cos^{-1}\left(\sqrt{\frac{I_s\left(\frac{M^2}{2E}-I_i\right)}{\frac{M^2}{2E}(I_s-I_i)}}\right). \quad (A70)$$

When $M^2/2E \to I_s$, $\theta_S \to 0$ and when $M^2/2E \to I_i$, $0 < \theta_S < \pi/2$. Note that this behavior is analogous to the behavior of $\theta_L$ in L-convention in the case of a LAM (cf. equations (A39) and (A40) in SA91). If $I_i = I_l$, then $\theta_S$ is constant.

From equation (A68), we can see that the minimum and maximum values of the rate of change of $\phi_S$ will occur when $\operatorname{sn}\tau = \pm 1$ and $\operatorname{sn}\tau = 0$.
I.e.,
$$(\dot{\phi}_S)_{min} = \frac{M}{I_i}, \quad (A71)$$

and

$$(\dot{\phi}_S)_{max} = \frac{M}{I_l}. \quad (A72)$$

The angle $\psi_S$ is periodic with a period $P_{\psi_S}$ and is given by (cf. Landau and Lifshitz 1976, pp. 118-119)

$$P_{\psi_S} = 4\sqrt{\frac{I_l I_i I_s}{2E(I_s - I_i)\left(\frac{M^2}{2E} - I_l\right)}} \int_0^{\pi/2} \frac{du}{\sqrt{1 - k^2 \sin^2 u}} \tag{A73}$$

and is same as the period of sn $\tau$. Furthermore, this period of the angle $\psi_S$ is same as the period of the angle $\psi_L$ in the L-convention for SAMs (compare equation (A73) with equation (A71) in SA91).

Therefore, for SAMs, as for LAMs (see equation (A48)),

$$P_{\psi_S} = P_{\psi_L}. \tag{A74}$$

From equation (A73), we see that the period of $\psi_S$ will have minimum and maximum values when $k^2 = 0$ and $k^2 = 1$ respectively.

When $M^2/2E = I_s$, $k^2 = 0$ and

$$P_{\psi_S} = 2\pi \sqrt{\frac{I_l I_i I_s}{2E(I_s - I_i)(I_s - I_l)}}.$$

i.e.,

$$(P_{\psi_S})_{\min} = \frac{2\pi I_s}{M} \sqrt{\frac{I_l I_i}{(I_s - I_i)(I_s - I_l)}}. \tag{A75}$$

When $M^2/2E \rightarrow I_i$, $k^2 \rightarrow 1$ and the integral on the right side of equation (A73) $\rightarrow \infty$.
i.e.,

$$(P_{\psi_S})_{\max} \rightarrow \infty. \tag{A76}$$

Therefore for SAMs,

$$\frac{2\pi I_s}{M} \sqrt{\frac{I_l I_i}{(I_s - I_i)(I_s - I_l)}} \leq P_{\psi_S} < \infty \tag{A77}$$

From equation (A68), it can be seen that $\dot{\phi}_S$ is periodic with period $P_\psi/2$. The angle $\phi_S$ is aperiodic unless $L_i = L_l$ or $\int_0^{P_{\psi_S}/2} \dot{\phi}_S dt = 2\pi$ (this latter condition may not exist for certain values of axial lengths $L_l$, $L_i$, and $L_s$). Similar to LAMs, one can define a time-averaged period for $\phi_S$, $P_{\overline{\phi}_S}$, where $P_{\bar{\dot\phi}_S} = 2\pi/(\dot\phi_S)_{mean}$ and $(\dot\phi_S)_{mean} = (2/P_{\psi_S}) \int_0^{P_{\psi_S}/2} \dot\phi_S \, dt$. (Note: Again, for simplicity, we will denote $P_{\bar{\dot\phi}_S}$ simply by $P_{\phi_S}$). Based on the properties of sn $\tau$, from equation (A68) we can deduce that $P_{\phi_S}$ is minimum when $k^2 \to 0$ (i.e., when sn $\tau \to$ sin $\tau$ and $M^2/2E \to I_s$) and maximum when $k^2 \to 1$ (i.e., when sn $\tau \to$ tanh $\tau$ and $M^2/2E \to I_i$).

Therefore from equation (A68), we have

$$\left(P_{\phi_S}\right)_{min} = \frac{2\pi I_s \sqrt{I_l I_i}}{M\left(\sqrt{I_l I_i} + \sqrt{(I_s - I_i)(I_s - I_l)}\right)} \tag{A78}$$

and

$$\left(P_{\phi_S}\right)_{max} = \frac{2\pi I_i}{M}. \tag{A79}$$

Therefore for SAMs,

$$\frac{2\pi I_s \sqrt{I_l I_i}}{M\left(\sqrt{I_l I_i} + \sqrt{(I_s - I_i)(I_s - I_l)}\right)} \le P_{\phi_S} \le \frac{2\pi I_i}{M}. \tag{A80}$$

Note that in the L-convention (unlike $(P_{\phi_S})_{min}$ and $(P_{\phi_S})_{max}$ in the S-convention), for SAMs, $(P_{\phi_L})_{min}$ occurs when $M^2/2E \to I_i$ (i.e., $k^2 \to 1$) and $(P_{\phi_L})_{max}$ occurs when $M^2/2E \to I_s$ (i.e., when $k^2 \to 0$) (cf. SA91).

Still in the S-convention, $\left(\dfrac{P_{\psi_S}}{P_{\phi_S}}\right)_{min}$ occurs when $k^2 \to 0$ and $\left(\dfrac{P_{\psi_S}}{P_{\phi_S}}\right)_{max}$ occurs when $k^2 \to 1$.

From equations (A75) and (A78), we obtain

$$\left(\frac{P_{\psi_S}}{P_{\phi_S}}\right)_{min} = \frac{\sqrt{I_l I_i} + \sqrt{(I_s - I_i)(I_s - I_l)}}{\sqrt{(I_s - I_i)(I_s - I_l)}} \tag{A81}$$

whereas from equations (A76) and (A79), we have

$$\left(\frac{P_{\psi_S}}{P_{\phi_S}}\right)_{max} \to \infty. \tag{A82}$$

i.e.,

$$\left(\frac{P_{\psi_s}}{P_{\phi_s}}\right) \geq \frac{\sqrt{I_l I_i} + \sqrt{(I_s - I_i)(I_s - I_l)}}{\sqrt{(I_s - I_i)(I_s - I_l)}} \quad . \tag{A83}$$

Using equations (A27), (A28), and (A29) we can rewrite equation (A83) as

$$\left(\frac{P_{\psi_s}}{P_{\phi_s}}\right) \geq 1 + \sqrt{\frac{(L_i^2 + L_s^2)(L_l^2 + L_s^2)}{(L_i^2 - L_s^2)(L_l^2 - L_s^2)}} \quad . \tag{A84}$$

Since $\sqrt{\frac{(L_i^2 + L_s^2)(L_l^2 + L_s^2)}{(L_i^2 - L_s^2)(L_l^2 - L_s^2)}} > 1$,

$$\left(\frac{P_{\psi_s}}{P_{\phi_s}}\right) > 2. \tag{A85}$$